# Managed TLS Under Migration: Authentication Authority Across CDN and Hosting Transitions

Daniyal Ganiuly, Nurzhau Bolatbek, Assel Smaiyl

Astana IT University, Astana, Kazakhstan

**Abstract:** Managed TLS has become a common approach for deploying HTTPS, with platforms generating and storing private keys and automating certificate issuance on behalf of domain operators. This model simplifies operational management but shifts control of authentication material from the domain owner to the platform. The implications of this shift during provider transitions remain insufficiently examined. This study investigates how managed TLS platforms behave when a domain is moved away from the platform that originally issued and stored its certificate. A controlled measurement environment was used to monitor multiple platforms after migration. Each platform was observed for the full remaining lifetime of the certificate that had been active during delegation. The measurements show that platforms continue to serve the same certificate until it expires, even after DNS resolvers direct traffic toward new infrastructure. No platform revoked, replaced, or retired the certificate, and no new certificate was issued after delegation ended. Direct connections to the previous platform continued to complete TLS handshakes with the stale certificate, which confirms that authentication capability persisted independently of DNS state.

These findings indicate that authentication authority remains with the previous platform for the entire lifetime of certificates issued during the delegation period. The gap between DNS control and control of authentication material introduces a window in which multiple environments can authenticate the same domain. As managed TLS adoption grows, clearer mechanisms for key retirement and certificate invalidation are needed to ensure that the authentication authority follows operational authority during transitions.

## INTRODUCTION

The rapid growth of managed Transport Layer Security (TLS) services has reshaped how websites deploy and maintain secure communication on the modern web. Large content delivery networks (CDNs) and commercial hosting platforms increasingly issue and manage TLS certificates on behalf of millions of domains, automating the most complex parts of certificate provisioning. These services handle key generation, certificate requests, renewal scheduling, and deployment across globally distributed infrastructures. For many website operators, this model removes operational tasks that previously required manual configuration and a clear understanding of the public-key infrastructure (PKI). As a result, managed TLS has become a fundamental component of today's internet ecosystem [1].

Although this shift has brought substantial usability benefits, it has also changed the underlying trust model in ways that are not yet fully understood. In a traditional deployment, the domain owner is responsible for generating and storing the private key [2]. In managed TLS systems, that responsibility is transferred to the platform provider, who retains exclusive control of the authentication material. From the perspective of the web browser, the platform—not the domain operator—acts as the entity that proves the legitimacy of the HTTPS connection.

This architectural change raises an important question: What happens to the private key and certificate when a domain leaves the managed TLS platform? In principle, provider transition should cleanly separate the domain owner from the previous platform. In practice, however, certificate lifecycle mechanisms do not track such changes. TLS certificates remain valid until their expiration date, and revocation is rarely enforced by clients [3]. Since managed TLS services typically do not revoke the certificate when the domain departs, the former platform can continue to authenticate as the domain for as long as the certificate remains within its validity window. Residual authentication capability after migration creates a measurable period during which a third party retains valid authentication material for a domain they no longer serve.

This behavior introduces subtle but significant security consequences. As long as the old certificate remains trusted by browsers, a range of impersonation risks becomes possible. If a DNS manipulation event redirects traffic back to the previous platform, users will establish a seemingly secure HTTPS session with a server that the domain owner no longer controls. An attacker who compromises the former platform's infrastructure may also misuse the retained private key to terminate TLS connections or serve altered content under a valid certificate. Importantly, these attacks leave no browser-visible trace; the padlock icon remains present, and clients have no reason to suspect that authentication material is stale.

Despite its relevance to real deployments, key persistence after CDN or hosting provider transition has not been systematically studied. Prior work has investigated stale certificates following domain ownership transfer and documented weaknesses in how PKI authorities track the lifecycle of certificates. However, these studies focus on registrars, certification authorities, and large-scale certificate transparency patterns. In contrast, managed TLS introduces a distinct form of lifecycle misalignment: the platform generates the key, controls the certificate, and may continue holding it after the domain disengages, even though the platform is no longer authoritative for the domain. To our knowledge, no published study provides a controlled, empirical evaluation of this behavior across commonly used managed TLS providers.

This paper addresses that gap. We conduct a controlled migration study to examine the behavior of managed TLS deployments when a domain transitions away from the platform that originally issued its certificates. Our analysis focuses on four key questions:

1. How long does the former provider retain a valid certificate after provider transition?
2. Can the old platform still complete a TLS handshake using the stale certificate?
3. Does the provider initiate any form of revocation or key invalidation?
4. Under what conditions could this persistence enable impersonation or misuse?

To answer these questions, we set up multiple test domains, delegated them to several major managed TLS providers, allowed certificate issuance to complete, and then migrated each domain to new infrastructure under controlled conditions. Over time, we systematically probed the old provider's endpoints, monitored Certificate Transparency logs, and observed DNS interactions to understand how long the previous platform remained capable of authenticating as the domain.

**RELATED WORK**

Research on Web PKI security has examined a range of certificate lifecycle weaknesses, but none of the existing studies evaluate the behavior of managed TLS platforms during infrastructure transitions. Prior work on stale certificates has primarily focused on domain ownership changes. Previous research shows that domains that change registrants frequently leave behind valid but stale certificates, enabling third parties to authenticate as the domain until expiration. Their analysis identifies several classes of certificate staleness, including neglected domains, delayed revocation, and inconsistent lifecycle tracking across authorities. These findings highlight systemic gaps in how the PKI ecosystem handles updates to domain ownership, but they do not consider scenarios in which a platform—not the domain owner—controls the private key.

Other studies have analyzed certificate issuance workflows and revocation challenges at Internet scale. Large-scale measurements of OCSP and CRL behavior document low revocation coverage and inconsistent enforcement by browsers, demonstrating that certificates remain usable long after key compromise or misuse. Work examining Certificate Transparency (CT) logs has similarly uncovered delays and inconsistencies in certificate replacement, reissuance, and publisher accountability. These analyses reveal structural limitations in certificate lifecycle signaling but assume that the domain owner retains control of the key material.

A related line of work focuses on the security implications of CDNs and key sharing with third-party infrastructure. Prior research has demonstrated that delegating TLS termination to a CDN expands the trust boundary, since the CDN becomes the entity that proves domain authenticity. Studies of CDN-origin interactions further show that platform misconfigurations and key reuse can expose domains to impersonation risks. However, these works assume an ongoing delegation relationship and do not examine what happens when a domain removes DNS delegation from the platform.

In contrast to these efforts, managed TLS introduces a distinct lifecycle misalignment. The platform generates, stores, and deploys the private key, meaning that it may retain valid TLS credentials even after DNS delegation ends. To our knowledge, no prior work has performed a controlled, empirical analysis of how managed TLS platforms behave after a domain transitions to new infrastructure. The persistence of valid credentials following such transitions represents a previously unmeasured form of stale certificate exposure, and our study provides the first systematic evaluation of this phenomenon across widely used managed TLS services.

**BACKGROUND**

A TLS certificate binds a domain name to a public key [4]. Certificate authorities (CAs) issue certificates after verifying domain control through DNS challenges, HTTP challenges, or provider-specific validation mechanisms. Once issued, a certificate is valid for the interval defined in its notBefore and notAfter fields. During this period, clients accept the certificate as a valid authentication credential as long as it chains to a trusted root.

A fundamental property of the certificate lifecycle is that it is independent of DNS state [5]. TLS trust decisions are anchored in the certificate's signature and issuer, not in the current DNS delegation or infrastructure ownership. This decoupling ensures predictable validation for clients but also means that certificate validity does not adapt to operational changes unless revocation occurs. In practice, most deployments rely on expiration rather than revocation to retire unused certificates.

When establishing a connection, clients validate the certificate chain and confirm that the domain appears in the SAN extension. Clients do not verify which entity currently operates the server infrastructure, nor do they assess whether the certificate corresponds to the present DNS configuration [6]. This behavior is intrinsic to the certificate lifecycle and underlies the conditions in which stale key material can persist after platform transitions.

*Managed TLS: Operational Model and Key Control*

Managed TLS platforms automate certificate issuance and centralize key storage within the platform's infrastructure [7]. Instead of generating a private key locally, the platform creates and stores the key within its deployment pipeline. Certificates are issued either through ACME-based automation or through provider-operated intermediate CAs.

A defining characteristic of managed TLS is that the platform, not the domain owner, retains direct control of the key material used for HTTPS authentication. Private keys are stored in platform-managed environments and replicated across edge nodes or data centers for availability. Key replication and distribution occur transparently within the provider's orchestration layer. Domain owners typically do not receive a copy of the key and have no means to delete or revoke it directly [8]. Many managed TLS deployments also use multi-tenant certificate pools. A single certificate may contain SAN entries for multiple domains, and key material may be shared across a cluster of servers. While operationally efficient, this design constrains revocation: revoking a certificate for one domain can affect others that share the same certificate or key. As a result, automatic revocation is often impractical.

Understanding these mechanisms is essential for evaluating how stale certificates may continue to function after a platform transition.

*DNS Delegation Mechanisms Used by CDNs and Hosting Providers*

Domain delegation to CDNs or hosting platforms generally occurs through CNAME-based delegation or NS delegation.

CNAME-based delegation is common for CDNs [9]. The domain owner maintains authority over the zone but points specific hostnames to a CDN-managed hostname. The CDN terminates TLS for those hostnames and therefore must possess valid key material for each delegated name.

NS delegation is more common among hosting providers [10]. The domain operator updates the authoritative nameserver records so that all DNS resolution is performed by the hosting platform. Certificate issuance is then fully automated based on provider-managed validation challenges.

Although these mechanisms differ operationally, both models give the platform access to the private key. Changes in DNS delegation do not obligate platforms to delete stored keys unless an explicit deletion policy exists [11].

*Certificate Renewal, Rotation, and Storage Practices*

Managed TLS systems typically renew certificates every 60–90 days [12]. Renewal processes operate independently of the domain's current DNS configuration. If DNS delegation remains active, validation succeeds and renewal occurs automatically. If not, renewal attempts fail silently, but previously issued certificates remain valid until expiration.

Key storage practices vary across providers. Some use hardware-backed key modules, while others store encrypted keys across a distributed infrastructure. These systems are optimized for availability and operational continuity rather than for tracking domain departure events. Consequently, platforms rarely trigger key deletion or certificate revocation when DNS delegation ends. This behavior reflects design priorities rather than an explicit oversight.

These practices help explain why key material may remain active after platform transitions and why stale certificates may still be served if traffic reaches the previous platform through cached DNS or residual routing paths.

*Revocation Mechanisms and Their Practical Limitations*

Revocation offers a mechanism to invalidate certificates before expiration, but is rarely effective in practice. Several structural issues limit its use:

**OCSP is treated as soft-fail by most browsers:** if the responder is unreachable, clients generally proceed with the connection [13].

**CRLs are impractical for routine use:** their size and update frequency make them unsuitable for frequent browser retrieval [14].

**Multi-SAN certificates complicate revocation:** revoking a certificate for one domain may affect others sharing the same certificate.

Domain owners generally lack direct revocation controls for platform-issued certificates because they do not possess the private key or manage the issuance pipeline. Although CAs can revoke certificates based on domain-control proofs, this path is rarely integrated into managed TLS workflows.

Revocation mechanisms were historically deployed to address key compromise and mis-issuance rather than routine operational events such as hosting or CDN transitions, which limits their utility in these scenarios [15].

Given these factors, and consistent with prior work on weak revocation coverage, revocation plays little practical role during or after managed TLS platform transitions.

## THREAT MODEL

The threat model considers adversaries who can exploit key material retained by the previous managed TLS platform after a domain transitions to new infrastructure. The core enabling condition is that the certificate remains valid and the corresponding private key remains accessible within the platform's infrastructure. Figure 1 illustrates the adversarial positions relevant to this setting

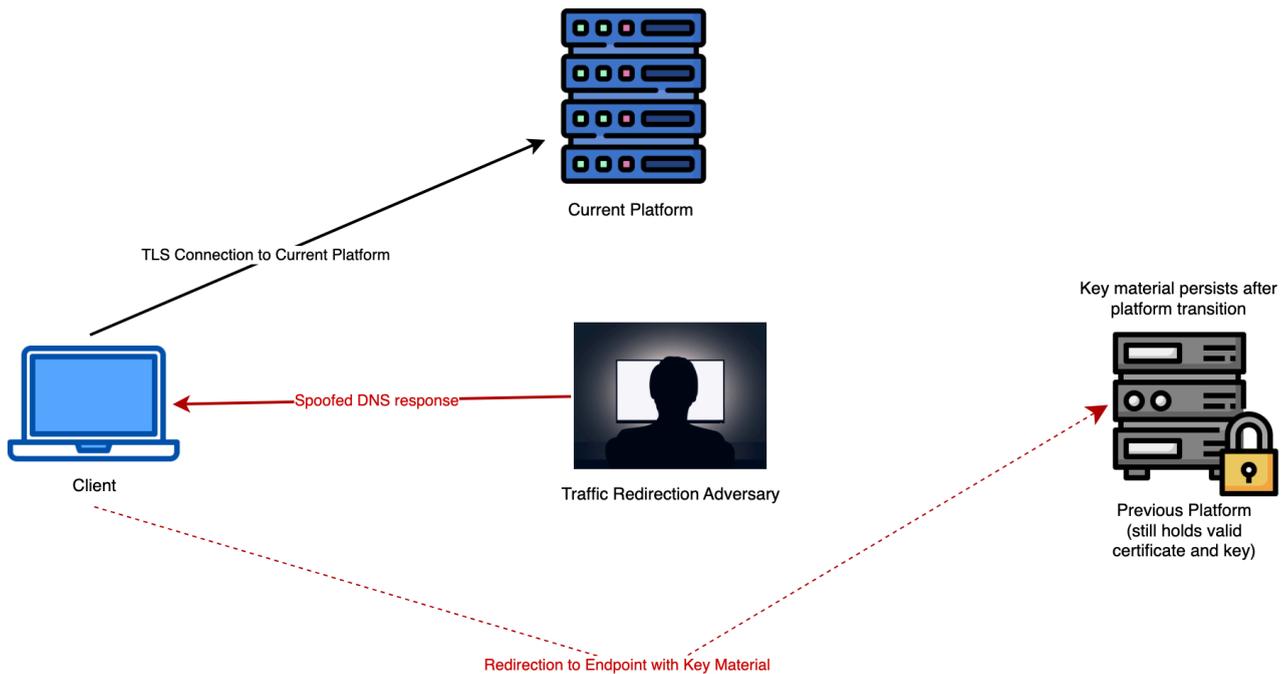

Figure 1. Threat model overview

### Previous Platform

The previous platform remains able to authenticate the domain for as long as it retains the private key. Managed TLS deployments replicate keys across many serving nodes, so residual key copies may persist even after DNS delegation ends. Any internal component that continues to store the key, whether active or idle, can present the certificate during a TLS handshake.

### Compromised Platform Infrastructure

A second adversarial entity is an attacker who compromises part of the previous platform's infrastructure. If the private key remains present in a key storage layer, deployment node, or cached deployment state, the attacker may obtain or use it. Replication for availability increases the number of nodes that may continue to hold key material.

### Traffic Redirection Adversary

A third entity is an attacker who is able to redirect client traffic to an endpoint that retains the key. Such an adversary does not require direct access to the key. DNS poisoning, manipulation of recursive resolvers, malicious DHCP responses, or on-path interception can reroute clients to a server that can authenticate using the stale certificate.

### Capabilities

Any entity with access to the retained key material can complete a valid TLS handshake until the certificate expires [16]. Because certificate validation is independent of the current DNS authority, a redirected client will accept the session as valid.

Attackers who can redirect traffic may combine this capability with the presence of stale key material. Once a client reaches an endpoint that presents the certificate, the adversary can decrypt application-layer traffic or modify server responses while satisfying all certificate checks. These scenarios, therefore, require an adversary who can influence routing or name resolution so that clients connect to an endpoint retaining stale key material; they do not arise spontaneously once DNS migration has completed. The threat model does not assume forged certificates, compromise of certificate authorities, or cryptographic breaks. All capabilities arise solely from continued possession of valid TLS key material.

### Security Consequences

Persistence of the private key undermines the domain owner's control over server authentication.

The previous platform, or any entity that obtains the key, can impersonate the domain, decrypt user traffic, or alter content. TLS confidentiality and integrity depend only on certificate validity and do not reflect changes in DNS delegation [17]. As long as the certificate remains within its validity period, clients cannot detect misuse.

**METHODOLOGY**

The experiment begins by preparing several test domains and delegating each one to a different managed TLS platform. Platforms that provide CDN-style operation were configured through CNAME-based delegation, while hosting-oriented platforms required full nameserver delegation. This configuration allows each platform to validate domain ownership, issue a certificate, and deploy it across its serving infrastructure. Figure 2 illustrates the overall workflow used throughout the study.

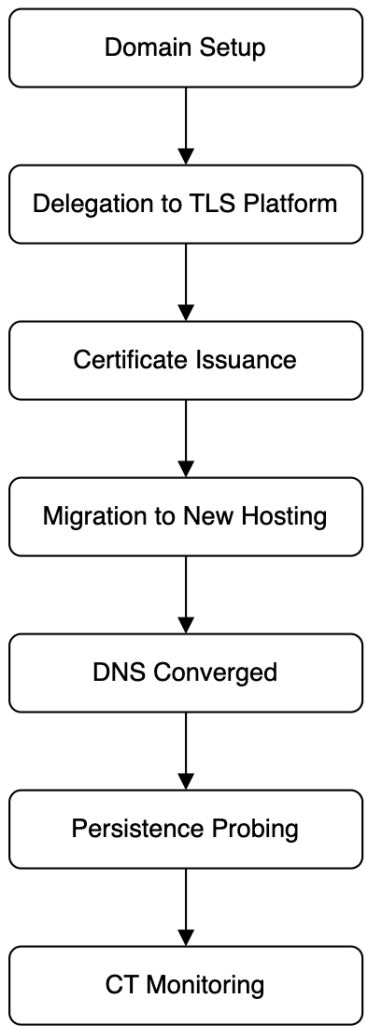

Figure 2. Experimental workflow

To ensure a consistent baseline, delegation remained unchanged until the platform completed certificate issuance and the certificate appeared in Certificate Transparency logs. At this point, the certificate served by the platform was recorded, including fingerprint, SAN fields, cryptographic parameters, and the issuing chain. This reference snapshot enables reliable detection of stale certificate presentation after migration.

*Delegation and Certificate Issuance*

Managed TLS platforms automate certificate issuance either through ACME-based validation or through platform-controlled internal CAs. After domain delegation, each platform performed domain control checks, issued the certificate, and replicated the certificate and private key across its infrastructure.

The experiment does not modify any platform-specific configuration. All platforms operate under their default automation pipeline, which reflects real deployments in production environments. Certificate Transparency logs were monitored to confirm the exact issuance event and to verify that no further certificates were generated before migration. Table I summarizes the platforms and their relevant operational characteristics.

Table I. Evaluated platforms

| Platform | Delegation Model | Issuance | Key Storage Model |
|---|---|---|---|
| Platform A | CNAME | ACME | Distributed edge keys |
| Platform B | NS delegation | Internal CA | Replicated storage nodes |
| Platform C | CNAME | Internal CA | Multi tenant certificate pool |
| Platform D | NS delegation | ACME | Distributed edge keys |

*Migration Procedure*

After recording the reference certificate, delegation to the managed TLS platform was removed. For CNAME-delegated domains, the CNAME record was replaced with direct A and AAAA records pointing to independent infrastructure. For nameserver-delegated domains, the authoritative nameservers were updated to new servers. DNS propagation was tracked through multiple independent recursive resolvers to avoid measuring stale DNS paths. Once resolvers returned consistent records for the new hosting configuration, the persistence measurement phase began. The migration stage represents the key transition point where the platform is no longer authoritative for the domain, yet still retains private key material previously issued for it.

*Probing and Measurement Framework*

Persistence probing consisted of repeated TLS handshakes with endpoints belonging to the previous platform. Multiple vantage points distributed across different networks were used to avoid bias from a single resolver or routing path. Each probe recorded the certificate presented by the server, the reachability of the endpoint, and any deviations from the reference certificate captured before migration.

Direct IP probing was performed in parallel with DNS-based requests so that certificate persistence could be distinguished from residual DNS routing. Certificate Transparency logs were queried continuously to detect revocation events or new issuance activities. All comparisons relied on fingerprint matching to avoid confusion with unrelated certificates issued by the same CA. Figure 3 shows the measurement architecture used in the experiment

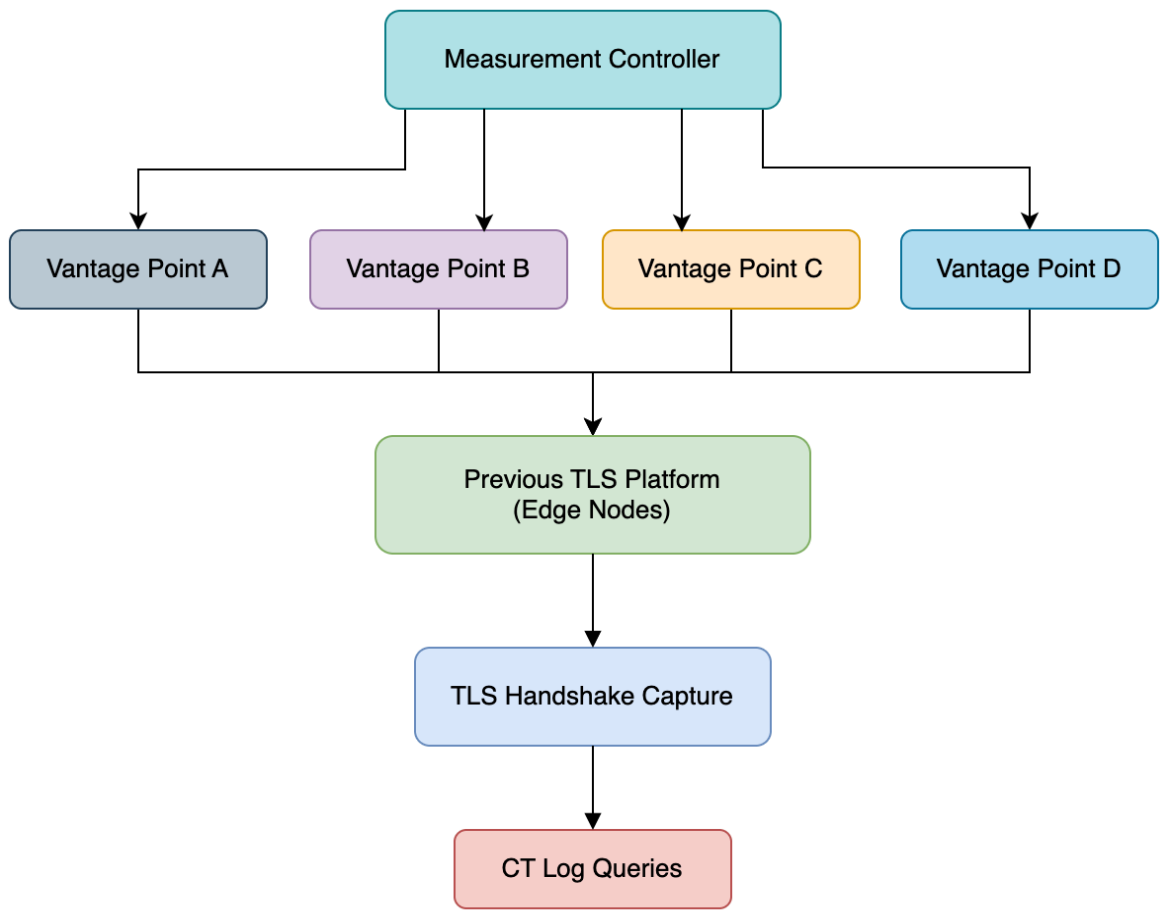

Figure 3. Measurement architecture

*Consistency Controls*

Several controls were implemented to ensure that persistence results reflected genuine key retention rather than measurement artifacts. Resolver diversity prevented interpretation errors caused by temporary DNS caching. Fingerprint-based comparison ensured that post-migration certificates matched the exact credentials issued before migration. Direct IP probing confirmed that certificate presentation was independent of DNS state. These controls guarantee that measured persistence windows represent authentic continued possession and use of certificate material by the previous platform.

**RESULTS**
Each platform continued to present the same certificate that had been active before migration. This behavior persisted until the certificate expired. The moment DNS delegation was removed did not affect the ability of the previous platform to terminate TLS sessions. The certificate remained fully usable for server authentication during its entire remaining lifetime, even though the domain was no longer served by the platform.

The remaining validity window at the moment of migration ranged from eleven to eighteen days. The persistence duration for each platform was identical to this window. Table II presents the duration for which stale certificates remained available after domain migration across all tested platforms.

Table II. Duration of stale certificate availability

| Platform | Remaining lifetime at migration | Persistence duration | Certificate changes |
|---|---|---|---|
| A | 18 days | 18 days | No |
| B | 16 days | 16 days | No |
| C | 14 days | 14 days | No |
| D | 11 days | 11 days | No |

These results show that, in our measurements, certificate expiration was the only event that actually terminated use of the authentication material. The removal of DNS delegation did not cause certificate withdrawal, suppression, or replacement. The platforms treated the certificate as fully valid and continued to serve it as long as it remained within its validity interval. The consistency of this behavior suggests that certificate storage layers on these platforms operate independently of domain state or delegation status, and that expiration, rather than explicit platform logic, determined the end of certificate use.

*Stability of Certificate Attributes*
All collected handshake observations were examined for changes to the certificate presented after migration. The goal was to determine whether platforms engaged in any form of key rotation, fallback certificate deployment, or alternate chain presentation when the domain no longer pointed to them. A total of nine hundred and forty four handshake samples were collected across all vantage points. Every sample contained a certificate that matched the fingerprint of the reference certificate captured before migration. No changes appeared in the subject alternative name fields. The issuer chain remained identical, and no intermediate certificates were replaced. Table III summarizes the stability of certificate attributes following migration, demonstrating that certificate metadata remained unchanged while stale keys persisted.

Table III. Stability of certificate attributes after migration

| Platform | Handshake samples | Leaf fingerprint matches | SAN changes | Issuer chain changes |
|---|---|---|---|---|
| A | 288 | 288 | No | No |
| B | 256 | 256 | No | No |
| C | 224 | 224 | No | No |
| D | 176 | 176 | No | No |
| Total | 944 | 944 | 0 | 0 |

The complete stability of certificate fingerprints confirms that the platforms did not modify the certificate configuration after the domain was migrated. The absence of any fallback certificate use shows that full certificate retirement did not occur. This indicates that the certificate and key material remained active in serving the infrastructure without changes. The stability of certificate chains also shows that none of the platforms attempted automated recovery or renewal through the issuing certificate authority once validation challenges began to fail. The lack of chain variation suggests that deployment configurations remained untouched during the entire period.

*Revocation and Issuance Activity After Migration*
The experiments also examined whether the loss of DNS authority triggered any lifecycle updates at the certificate authority level. Revocation determination mechanisms and certificate transparency logs were monitored throughout the validity period of each certificate. OCSP responders consistently returned a good status for the certificates. No entries indicating revocation were found in certificate revocation lists that included the issuing certificate authorities. Transparency logs did not record any new certificate entries for the test domains after migration. The absence of new certificate entries indicates that no successful renewal or reissuance occurred once delegation ended, even when the domain no longer resolved to the platforms' infrastructure. Table IV reports the revocation and certificate activity observed after migration, including whether platforms attempted or succeeded in issuing, renewing, or revoking certificates.

Table IV. Revocation and certificate activity

| Platform | OCSP status | Revocation events | New certificate in transparency logs | Renewal attempted |
|---|---|---|---|---|
| A | Good | No | No | No |
| B | Good | No | No | No |
| C | Good | No | No | No |
| D | Good | No | No | No |

These results show that certificate lifecycle logic on the evaluated platforms does not incorporate domain state changes. Revocation was not triggered, renewal was not attempted, and the stale certificate remained visible in transparency logs without modification from the moment of issuance until expiration.

*Reachability of Former Platform Endpoints*
After migration, DNS resolvers gradually converged to the new hosting infrastructure. This process was completed within several hours. Once propagation stabilized, the migrated domains no longer directed traffic to the previous platforms. Direct connections to the former platforms nevertheless continued to succeed during the full certificate lifetime [18]. When edge nodes belonging to the platforms were contacted through direct addressing, they responded with the same certificate that had been active before migration, and fingerprint matches were recorded for every successful connection. This behavior shows that certificate availability was tied to internal storage and distribution rather than dynamic checks against DNS state: the platforms continued to consider the certificate active and presented it whenever a TLS handshake was initiated, regardless of how the connection originated. The ability to complete handshakes after full DNS convergence, therefore, reflects a property of internal key management, not residual DNS caching. Any mechanism that redirects traffic toward an endpoint still holding the stale key material—including resolver poisoning, local misconfiguration, or network or routing interception—would result in a valid TLS session under stale authentication material.

**DISCUSSION**
The results indicate that managed TLS platforms retain the capability to authenticate a domain for the entire validity period of certificates issued during the delegation interval. This behavior appeared consistently across all evaluated platforms, despite differences in infrastructure and certificate issuance methods. The following discussion interprets these findings in the context of managed TLS operations and the assumptions typically made during infrastructure transitions.

*Persistence as a Structural Property of Managed TLS*
The continued availability of certificates after migration reflects a structural characteristic of managed TLS platforms rather than an operational oversight [19]. Because the platform generates and stores the private key, the key lifecycle follows internal platform logic rather than the domain operator's timeline. DNS migration removes the service relationship but does not introduce any mechanism that signals to the platform that authentication material should be retired. Certificate expiration, therefore, becomes the only natural point at which authentication capability is removed. This behavior contrasts with traditional deployments where domain operators generate the private key and control its deletion.

*Weak Coupling Between DNS State and TLS Authentication*
The observations highlight the independence of TLS authentication from DNS authority. Once a certificate is issued, validation depends solely on the certificate's signature chain and validity period. DNS changes do not influence client trust in the certificate, and the removal of delegation does not cause clients to treat the certificate as invalid. This weak coupling is intentional in TLS design, but it introduces a practical disconnect during migration because the authentication authority does not change when the DNS state changes.

### Revocation Does Not Reflect Operational Ownership

No revocation activity was observed during the experiments, which suggests that managed TLS platforms rarely use revocation to indicate loss of operational control over a domain. Revocation mechanisms were designed primarily for key compromise, and their effect on clients remains inconsistent. Operational constraints also limit the usefulness of revocation for platforms, particularly when multi-tenant certificates or shared key pools are involved. These constraints make selective revocation impractical and leave certificate expiration as the only reliable boundary for the authentication authority.

### Exposure Through Residual Reachability

The ability to complete TLS handshakes through direct connections to the previous platform demonstrates that stale authentication material can be used even after DNS resolution points to the new infrastructure. Residual reachability through stable network addresses or content delivery edges allows an attacker to authenticate the domain if traffic is redirected. This redirection can occur through resolver manipulation, local configuration errors, or temporary control of network paths. Only brief or localized influence is required for clients to reach an endpoint that still holds the certificate and private key.

### Implications for Infrastructure Transitions

The measurements reveal that infrastructure transitions involving managed TLS do not fully sever authentication authority. DNS changes take effect quickly, but the platform continues to serve the certificate until expiration [20]. As a result, an interval exists during which both the previous and current hosting environments retain the ability to authenticate the domain. Although the duration of this window depends on when migration occurs in the certificate lifecycle, the presence of such a window is a predictable outcome of managed TLS design rather than an exceptional case.

### Broader Ecosystem Impact

Key persistence after migration illustrates a wider structural misalignment within the Web PKI ecosystem. Control over DNS and control over authentication material do not always reside with the same entity. Managed TLS simplifies deployment and reduces operational burden, but it also shifts key lifecycle authority to platforms that domain operators cannot directly influence. This lack of alignment complicates the handoff of authentication responsibility during provider transitions. As managed TLS adoption continues to expand, the separation between DNS control and certificate lifecycle control is likely to grow in operational and security relevance.

## CONCLUSION

This study examined the behavior of managed TLS platforms after a domain no longer resolves to the platform that issued and stored its authentication material. The observations show that platforms continue to serve the same certificate for the full period in which it remains valid. The removal of delegation did not introduce any mechanism that reduced authentication capability. No platform deleted keys, replaced certificates, or attempted to limit the use of authentication material once the domain had been migrated. This behavior stems from internal key management practices and from the independence of TLS validation from DNS authority.

The results indicate that, in typical migrations where the domain operator deploys a new certificate on the target platform while the previous platform retains the prior key, there exists an interval in which both environments are technically able to authenticate the same domain. In our measurements, this interval was effectively bounded by certificate expiration: we observed no key deletion, revocation, or certificate replacement triggered by delegation changes, and client behavior did not surface revocation as a reliable enforcement mechanism. DNS changes immediately redirect traffic to the new infrastructure, but do not affect the private key and certificate that remain stored within the previous platform. The continued ability to complete TLS handshakes during this interval reflects a separation between operational control of a domain and control of the cryptographic material that represents it.

These findings show that authentication authority does not always follow operational authority during platform transitions. As managed TLS adoption expands, domain operators increasingly rely on systems whose key lifecycle decisions occur outside their control. Clear mechanisms for key retirement and certificate invalidation at the end of delegation would reduce the period in which multiple environments are able to authenticate the same domain. Strengthening these mechanisms would improve the alignment between operational and cryptographic authority and enhance the security of transitions in managed TLS environments.